\documentclass[twocolumn,showpacs,preprintnumbers,amsmath,amssymb]{revtex4}

\usepackage{graphicx}
\usepackage{dcolumn}
\usepackage{bm}

\begin{document}


\title{Resolved Sideband Emission of InAs/GaAs Quantum Dots Strained by Surface Acoustic Waves}

\author{M. Metcalfe$^{1}$, S. M. Carr$^{2}$, A. Muller$^1$, G. S. Solomon$^{1,2}$, and J. Lawall$^{2}$}
\address{
$^1$The Joint Quantum Institute, National Institute of Standards and Technology, Gaithersburg, MD 20899 and University of Maryland, College Park, MD 20742
$^2$National Institute of Standards and Technology, Gaithersburg, MD 20899
}

\date{\today}

\begin{abstract}
The dynamic response of InAs/GaAs self-assembled quantum dots (QDs) to strain is studied experimentally by periodically modulating the QDs with a surface acoustic wave and measuring the QD fluorescence with photoluminescence and resonant spectroscopy. When the acoustic frequency is larger than the QD linewidth, we resolve phonon sidebands in the QD fluorescence spectrum. Using a resonant pump laser, we have demonstrated optical frequency conversion via the dynamically modulated QD, which is the physical mechanism underlying laser sideband cooling a nanomechanical resonator by means of an embedded QD.
\end{abstract}

\pacs{PACS numbers here}

\maketitle

\begin{figure}[t]
\includegraphics[scale=1]{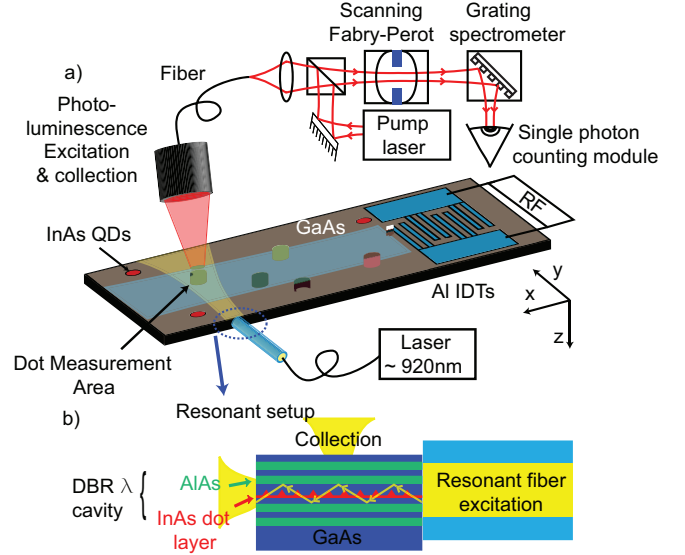}
\caption{\label{Chiplayout} (Color Online) {\bf (a)} Schematic of the experimental setup. {\bf (b)} Resonant fiber-waveguide coupling  (not to scale).
}
\end{figure}


Self-assembled InAs/GaAs quantum dots (QDs)
are sensitive optical probes of changes in their local environment. In particular, their discrete energy levels are sensitive to applied electric fields~\cite{Fry2000, Vogel2007} and to uniaxial \cite{Seidl2006B}, hydrostatic \cite{Itskevich1998, Lianga2006} and biaxial stresses \cite{Ding2010}.
Much of the focus to date has concerned probing the response of QDs to static perturbations; however, when perturbed dynamically at a rate exceeding the intrinsic QD linewidth, exciting new possibilities arise.
In the nanomechanical domain, for example, conventional optical probes become less effective as the device
size becomes smaller than the wavelength of light.
By employing a QD embedded in a nanomechanical beam as a microscopic sensor of strain,
laser sideband cooling a mechanical resonator to its quantum ground state
has been predicted to be possible in principle~\cite{Rae2004}.  Alternatively, it should be possible to
dynamically alter the fluorescence spectrum of a QD so as to generate entangled photon pairs~\cite{Gambetta2009}.

In this work, we characterize the dynamic response of embedded InAs/GaAs QDs by applying a periodic strain via a surface acoustic wave (SAW)~\cite{Matthews1977}. A SAW induces well-characterized, tunable strain components near a semiconductor surface at high frequencies.
We resolve strain-induced sidebands in QD fluorescence and demonstrate the physical basis of laser sideband cooling. We compare our experimental results to calculations based on static theory in which the response of QD level structure to strain is attributed to deformation potentials. While there is a rich history of using SAWs to modulate photonic
structures~\cite{Santos05,Sogawa01,Gell2008}, the work described here resides in a previously unexplored limit where the acoustic modulation frequency exceeds the resolvable optical linewidth (``resolved sideband limit'').

Our samples consist of InAs QDs embedded in a planar AlAs/GaAs distributed Bragg reflector (DBR) cavity
on which interdigitated transducers (IDTs) are fabricated for SAW generation (Fig.~\ref{Chiplayout}).
The cavity has fifteen DBR pairs below and ten pairs above the QD layer with a spacer optical thickness of 925~nm. The resulting cavity has a linewidth of $\approx350~\mathrm{GHz}$. This cavity enhances our collection efficiency by over an order of magnitude~\cite{benisty1998} and also enables us to perform resonant spectroscopy of our QDs (Fig.~\ref{Chiplayout}b).
The IDTs are are aligned so as to excite a SAW with a
cross-section of 30~$\mu$m propagating in the
[110] direction.  The IDT electrode period is $2.9~\mathrm{\mu m}$, corresponding to the wavelength $\lambda_s$ of a
SAW with a frequency $\nu_s$ of $1.05~\mathrm{GHz}$.

\begin{figure}[t]
\includegraphics[scale=1.0]{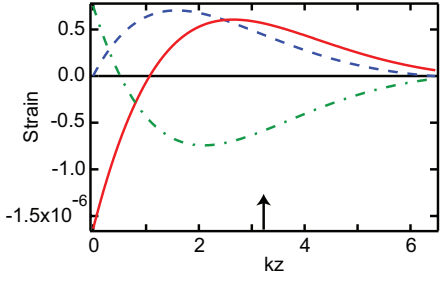}
\caption{\label{strains} (Color online) Amplitude of the SAW nonvanishing strains vs depth for a surface displacement of $1~\mathrm{pm}$ at a
frequency of $1~\mathrm{GHz}$.  $\varepsilon_{xx}$ (solid red) and $\varepsilon_{zz}$ (dot-dash green) are proportional to $\sin(kx-\omega_s t)$, while
$\varepsilon_{xz}$ (dash blue) is proportional to $\cos(kx-\omega_s t)$. Here $k=2\pi/\lambda_s$ and $\omega_s=2\pi\nu_s$. The black arrow corresponds to the sample parameters described in this paper: a SAW wavelength of $\lambda_s=2\pi/k=2.9~\mathrm{\mu m}$ and a QD depth of $z=1.5~\mathrm{\mu m}$.
}
\end{figure}
In order to quantify our experimental results and relate them to theoretical predictions, we relate the displacement and strain
fields~\cite{piezo,multilayer} induced by the SAW to the measurable surface displacement \cite{Simon1996, Santos05}.  We adopt a coordinate system in which the SAW propagates along the $\hat{x}$ direction, and the $\hat{z}$ direction is perpendicular to the surface and downwards (Fig.~\ref{Chiplayout}).
The vertical amplitude at the surface is proportional to the square root of the applied rf power
and is measured using a Michelson interferometer at room temperature. The displacement, $\vec{u}$, and strain, $\varepsilon_{ij}=(\partial_iu_j+\partial_ju_i)/2$, $i,j\in\{x,y,z\}$, can then be calculated at any depth within the sample.
As an example, the amplitudes of the inferred strains are plotted in Fig.~\ref{strains} as a function
of depth z for a surface displacement of 1 pm and a frequency of 1 GHz.

The sample is incorporated inside a homemade cryogenic microscope designed to allow for both photoluminescence (PL) and
resonant excitation spectroscopy at a sample temperature of $3~\mathrm{K}$.  QD fluorescence is collected by means of a fiber-coupled objective 
that can be scanned over the sample surface.  For PL, light from a laser with photon energy greater than the GaAs band gap is
injected into the collection optics in order to nonresonantly excite the QDs.
Alternatively, resonant excitation is performed by means of an optical fiber aligned to the edge of the cleaved chip (Fig.~\ref{Chiplayout}b), injecting quasimonochromatic laser (cavity-stabilized Ti:Sa, linewidth $<$~1~MHz) light into a waveguide mode of the DBR cavity~\cite{muller2007}. This guided mode inhibits the scattering of laser light into our collection optics. The resonant QD transition is driven by light in this mode, but its emission couples to a transverse FP mode of the planar cavity~\cite{benisty1998}. This light is then collected very efficiently perpendicular to the sample, without  scattered resonant pump light.
The fluorescence is analyzed by a homemade Fabry-Perot (FP) cavity (linewidth $250~\mathrm{MHz}$), whose length is scanned at a rate of $16~\mathrm{Hz}$. This is followed by a grating spectrometer to suppress transmission of all FP orders but one~\cite{Metcalfe2009}.
The spectrally filtered fluorescence is detected by means of a single-photon counting module, and photon arrival events
are correlated to the FP scan.

\begin{figure}
\includegraphics[scale=1.0]{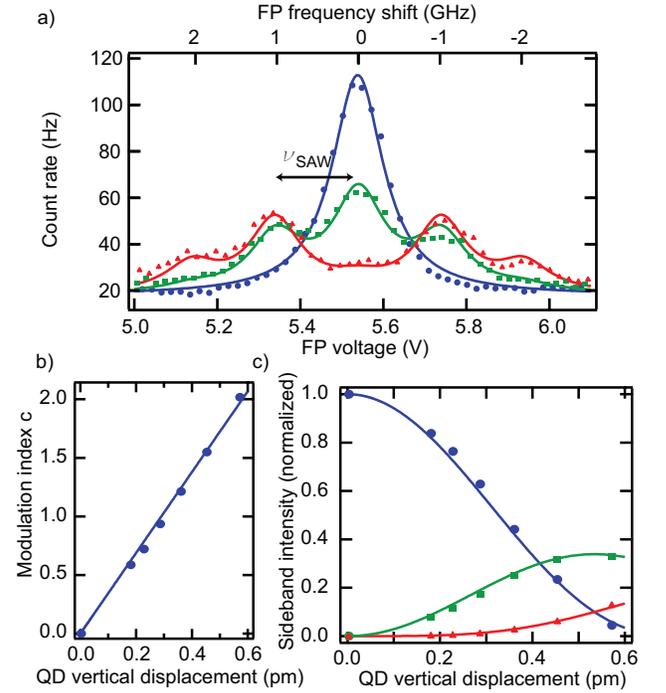}
\caption{\label{NarrowSAW} (Color online) {\bf (a)} QD fluorescence spectra at $921.5~\mathrm{nm}$ with no SAW excitation (blue circles), and SAW excitation characterized by QD displacements $|u_z|=1.2~\mathrm{pm}$ (green squares),
and $|u_z|=2.0~\mathrm{pm}$ (red triangles).  Lines are fits to equation (\ref{eqn: approxPspect}) to extract the modulation index $\chi$. Each FP spectrum is measured in $300~\mathrm{seconds}$.
{\bf (b)} Modulation index $vs$ $|u_z|$; $d\chi/d |u_z| \approx 3\times10^{12}~\mathrm{m}^{-1}$.
{\bf (c)} Peak intensities of the central QD peak (blue circles) and first- and second-order sidebands
(green squares, red triangles) vs QD displacement $|u_z|$. Lines are prediction of equation~(\ref{eqn: approxPspect}).
}
\end{figure}

Initially, individual QDs are located and studied by means of photoluminescence (Fig.~\ref{NarrowSAW}).
Once a QD with a bright and narrow (linewidth approximately $1~\mathrm{GHz}$) emission spectrum is located, the IDTs are driven at a frequency of
$\omega_s/2\pi=\nu_s=1.05~\mathrm{GHz}$, and the Lorentzian emission spectrum acquires sidebands spaced at the SAW frequency, $\nu_s$, as shown in Fig.~\ref{NarrowSAW}a.
As the rf power is increased, the central feature in the emission spectrum is depleted, and sidebands of higher order are generated.

To understand this behavior, we model the QD as a two-level system (TLS) with electric dipole operator $\hat{d}=d\sigma_x$, and dynamics
governed by the Hamiltonian
\begin{equation}
H=\frac{\hbar}{2}\left(\omega_0+\chi\omega_s \sin(\omega_s t)\right)\sigma_z
\label{eqn: TLS}
\end{equation}
and relaxation terms that cause the off-diagonal elements
of the density matrix $\rho$ to decay at a rate~$\gamma$.
Here the $\sigma_i$ are Pauli spin matrices, and the modulation index $\chi$ is a dimensionless
parameter expressing the frequency shift induced by the SAW on the TLS resonance frequency $\omega_0$ in units of $\omega_s$. The fluorescence is proportional to the expectation value $\langle \hat{d(t)}\rangle =Tr\left(\rho(t)\,\hat{d}\right)$.
Solving for the time evolution of the density matrix for weak incoherent excitation in the limit $\omega_s>>2\gamma$, the
power spectrum of the fluorescence is found to be proportional to $P[\omega]$, where
\begin{equation}
P[\omega]=\sum_{n=-\infty}^{\infty}\frac{J_n^2(\chi)}{\gamma^2+(\omega-(\omega_0-n\,\omega_s))^2}
\label{eqn: approxPspect}
\end{equation}
is a sum of Lorentzians with FWHM $2\gamma$ weighted by squares of Bessel functions, $J_n^2(\chi)$,
with maxima at frequencies $\omega_n=\omega_0-n\,\omega_s$, where $n$ is an integer.
In the limit $\gamma\rightarrow 0$ this is the well-known power spectrum of a frequency-modulated rf signal.

The solid lines in Fig.~\ref{NarrowSAW}a result from a fit to equation~(\ref{eqn: approxPspect}) in which the linewidth $2\gamma$,
central frequency $\omega_0$ and background are taken from the emission spectrum in the absence of SAW excitation, thus leaving $\chi$ as the only fitting parameter.
From a series of such curves, the height of each sideband, (normalized to the unmodulated peak), versus the calibrated SAW-induced displacement $|u_z|$ at the QD location, is shown in Fig.~\ref{NarrowSAW}c. Alternatively, Fig.~\ref{NarrowSAW}b shows the modulation index $\chi$ as a function of $|u_z|$.
The fit in Fig.~\ref{NarrowSAW}b reveals a displacement sensitivity of $d(\chi\nu_s)/d |u_z| \approx 3\times10^{21}~\mathrm{Hz/m}$,
and the solid lines in Fig.~\ref{NarrowSAW}c are the corresponding theoretical curves. Similar studies on four other QDs at this SAW frequency have found sensitivities in the range $3\times10^{21}~\mathrm{Hz/m} \pm 1.5\times10^{21}~\mathrm{Hz/m}$. We believe this variation is largely due to the fact that the SAW amplitude varies owing to its finite cross-section and we were unable to measure the SAW amplitude at the exact QD position $in\, situ$ in a cryogenic environment.

Sideband cooling a nanomechanical resonator with an embedded QD~\cite{Rae2004} involves
the quantized transfer of energy from a mechanical mode of the resonator
to an applied optical field.  To explore this matter we employ resonant spectroscopy (Fig.~\ref{Chiplayout}b). The coupling of the resonant laser to the TLS is described by adding to the Hamiltonian~(\ref{eqn: TLS}) an interaction term
\begin{equation}
H_{int}=-dE_0\cos(\omega_L t)\sigma_x,
\end{equation}
describing the dipole coupling of the QD to a laser field $E_0\cos(\omega_L t)$.
Calculating the time dependence of the atomic dipole moment in steady state, for weak excitation, the power spectrum of the fluorescence
is found proportional to
\begin{equation}
P[\omega]=\sum^{\infty}_{n=-\infty}\left|\sum_{k=-\infty}^{\infty}\frac{J_{n+k}(\chi) J_k(\chi)}{\gamma-i(\omega_L-\omega_0+k\omega_s)}\right|^2 \delta(\omega-\omega_L+n\omega_s)
\label{eqn: resPspect}
\end{equation}
This is of the form of a series of discrete lines at frequencies $\omega_n=\omega_L-n\omega_s$, spectrally separated
from the excitation frequency by multiples of the SAW frequency, that are resonantly enhanced when the excitation frequency matches the QD resonance or one of the SAW-induced sidebands.  Physically, the appearance of emission frequencies differing from the excitation frequency corresponds to
the transfer of mechanical energy to the light field.

\begin{figure}
\begin{center}
\includegraphics[scale=1.0]{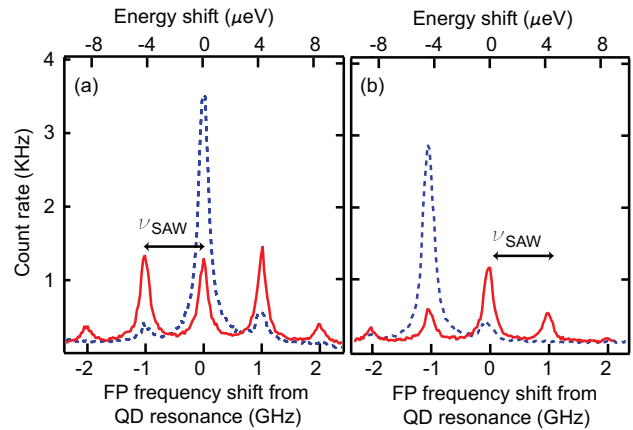}
\caption{\label{PLandRES} (Color online) The blue dashed curves correspond to a QD vertical displacement of $|u_z|=0.4~\mathrm{pm}$, and the red solid curves correspond to $|u_z|=1.0~\mathrm{pm}$. The QD emission wavelength is $924~\mathrm{nm}$. Each curve corresponds to $30~\mathrm{seconds}$ of data. (a) The QD is weakly driven on resonance and the spectrum of the resulting QD fluorescence is resolved with the scanning FP interferometer.
(b) The QD is driven by a laser tuned 1.05~GHz below the QD resonance (red sideband).  The fluorescence spectrum becomes asymmetric,
with average photon energy greater than the energy of the laser photons. }
\end{center}
\end{figure}

Figure~\ref{PLandRES}a shows the results of weakly driving the QD on resonance~\cite{auxiliary} for two different applied SAW powers. This particular QD has a linewidth of 1 GHz when measured in PL.  When resonantly driving weakly with narrowband
laser light, however, the fluorescence lines are quasimonochromatic~\cite{Loudon}, broadened
only by our FP analysis cavity.  The spectrum is thus much better resolved than in PL. The blue curves correspond to low SAW power and the spectra are dominated by re-emission at the pump frequency. High SAW powers are given by the red curves.  It is clear that more fluorescence power is frequency-shifted under the higher level of SAW excitation.
As expected, the separation between each peak in the spectrum is given by the SAW frequency ($1.05~\mathrm{GHz}$).
  Fitting a series of spectra such as those
shown in Fig.~\ref{PLandRES}a to the functional form given in equation~(\ref{eqn: resPspect}), we extract a displacement sensitivity agreeing closely with our previous value from PL measurements.  The fluorescence spectrum resulting when the QD is driven at a frequency $\nu_s$ below resonance
(red sideband) is shown in Fig.~\ref{PLandRES}b.  The asymmetry in the spectrum at high SAW power reflects the fact that on average
the energy of a scattered photon is larger than that of an incident photon, as mechanical energy is extracted from the SAW.
This is the physical basis of red sideband cooling. Similarly, when driving the QD at a frequency $\nu_s$ above resonance, the mean
photon emission energy is lower than that of the incident photons (data not shown). We have also studied QDs modulated at SAW frequencies of $90~\mathrm{MHz}$, $527~\mathrm{MHz}$ and $1585~\mathrm{MHz}$. The sideband separation is always equal to the SAW frequency and the sidebands disappear when we spectrally or spatially detune away from the QD.

To verify our interpretation of the experimental results, we neglect confinement effects \cite{Seidl2006B, Lianga2006}
and compare them to Pikus-Bir theory \cite{Harrison2005}, which
describes the dependence of the conduction and valence bands on applied static strain.
Neglecting terms quadratic in strain, and using our
SAW properties $\varepsilon_{yy}=\varepsilon_{xy}=\varepsilon_{yz}=0$, we estimate the valence band energy level shift to be
$\Delta E_v\approx2\pi\hbar(\varepsilon_{xx}(a_v+\frac{1}{2}b)+\varepsilon_{zz}(a_v-b))$,
where $a_v=2.4\times10^{14}~\mathrm{Hz}$ is the hydrostatic deformation potential of the valence band and $b=4.3\times10^{14}~\mathrm{Hz}$ is the principal shear deformation potential for GaAs. There is no dependence of the valence band shift on the SAW induced shear strain to first order.
In our experiment the QD is placed at a depth where $\varepsilon_{xx}\approx-\varepsilon_{zz}$ (Fig.~\ref{strains}), so the conduction band shift is negligible and the valence band shift reduces to $\Delta (\chi\nu_s)\approx\frac{3}{2}b\varepsilon_{xx}$ or $\Delta (\chi\nu_s)/\varepsilon_{xx}\approx\frac{3}{2}b=6.5\times10^{14}~\mathrm{Hz}$.  Expressing our experimental results in terms of
the SAW strain component $\varepsilon_{xx}$ yields $d(\chi\nu_s)/d\varepsilon_{xx}=1.3\times10^{15}~\mathrm{Hz} \pm 0.7\times10^{15}~\mathrm{Hz}$.
The level of agreement provides confirmation that the static theory is a reasonable approximation in the dynamic regime up to at least 1~GHz, but
the uncertainties do not allow detailed quantitative comparisons. For such purposes, a sample lacking an upper DBR stack and a cryogenic SAW calibration would be required.

By measuring the QD energy level shift, $\chi\nu_s$, at two different SAW frequencies and again assuming the effect of shear strain to be negligible, we are able to independently extract $d(\chi\nu_s)/d\varepsilon_{xx}$ and $d(\chi\nu_s)/d\varepsilon_{zz}$. We can then apply this calibration to estimate the frequency shift induced in a QD arising from motion in a nanomechanical beam. For a doubly-clamped beam fabricated in the $[110]$ direction, the dominant strain components are $\varepsilon_{xx}$ and $\varepsilon_{zz}=-0.45\varepsilon_{xx}$.
For the beam geometry chosen by Wilson-Rae~\cite{Rae2004}, with dimensions $950~\mathrm{nm} \times 30~\mathrm{nm} \times 85~\mathrm{nm}$,
the modulation index corresponding to thermal excitation at $4~\mathrm{K}$ is $\chi\approx 4$, so multiple sidebands would be observed.


In conclusion, we have studied the behavior of self assembled GaAs/InAs quantum dots under the application of surface acoustic waves. The SAW enables us to apply well controlled and tunable strain components to our QDs. We have demonstrated a QD displacement sensitivity
in the range $3\times10^{21}~\mathrm{Hz/m}\pm 1.5\times10^{21}~\mathrm{Hz/m}$
at $1.05~\mathrm{GHz}$. Phonon sidebands were resolved with both PL and resonant spectroscopy.
When driving the QD on the red sideband, the fluorescence spectrum was demonstrated to
consist of photons with mean energy greater than that of the incident photons, corresponding
to the extraction of mechanical energy from the SAW. In addition to applications in nanomechanics, this work offers a new avenue to pursue a deeper understanding
of the relation between QD energy level structure and applied strain.
It is clear from Fig.~\ref{strains} that the SAW offers the possibility
to explore the sensitivity of a QD to different strain configurations, depending on where the QD is located.
Another promising application of our SAW/QD system with resolved sidebands involves
using the biexciton cascade in InAs QDs as a source of entangled photon pairs ~\cite{Gambetta2009}.

The authors would like to thank G. Bryant, A. Chijioke, E. B. Flagg, W. Fang, K. Ekinci, M. Zaghloul and H.C. Ou. We acknowledge support by NSF though the Physics Frontier Center at JQI and the Center for Nanoscale Science and Technology for fabrication assistance.


\begin{thebibliography}{22}
\bibitem[{mul()}]{multilayer}
\bibinfo{note}{Here we neglect the multilayer structure of the sample due to
  the similar acoustic properties of AlAs and GaAs.}
\bibitem[{aux()}]{auxiliary}
\bibinfo{note}{The resonance fluorescence signal, along with the sidebands, are only measurable when a
 very small amount of incoherent excitation at 790~nm, far too
  low to produce an appreciable PL signal, is employed. An understanding of this phenomenon is under investigation}.
\end{thebibliography}
\end{document}